\documentclass[preprint,12pt,authoryear]{elsarticle}

\usepackage{amssymb}
\usepackage{booktabs}
\usepackage[table,dvipsnames]{xcolor}
\usepackage{colortbl}
\usepackage{amsmath}
\usepackage{lineno}
\usepackage{lipsum}
\usepackage{longtable,array,ragged2e}
\usepackage{multirow}
\usepackage{geometry}
\usepackage[version=4]{mhchem}
\usepackage{url}

\begin{document}

\begin{frontmatter}



\title{Explainable Multimodal Deep Learning Integrating Imaging and Clinical Data for Oral Potentially Malignant Disorder Detection} 

 \author[label1]{Ruilin You}
 \author[label1]{Yihan Wang}
 \author[label1]{Jiabin Chen}
 \author[label2]{Cherie Wink}
 \author[label2]{Petra Wilder-Smith}
 \author[label1]{Rongguang Liang}
 \author[label1]{Bofan Song\corref{cor1}}
 \affiliation[label1]{organization={Wyant College of Optical Sciences, University of Arizona},
             city={Tucson},
             state={AZ},
             country={USA}}

 \affiliation[label2]{organization={Beckman Laser Institute \& Medical Clinic, University of California Irvine},
             city={Irvine},
             state={CA},
             country={USA}}

\cortext[cor1]{Corresponding Author.}
\cortext[0]{E-mail addresses: \url{ruilinyou@arizona.edu} (R. You),
\url{jiabinchen@arizona.edu} (J. Chen),
\url{wangyihan@arizona.edu} (Y. Wang),
\url{zhihanhong@arizona.edu} (Z. Hong),
\url{cwink@hs.uci.edu} (C. Wink),
\url{pwsmith@hs.uci.edu} (P. Wilder-Smith),
\url{rliang@optics.arizona.edu} (R. Liang), and
\url{songb@arizona.edu} (B. Song).}
\begin{abstract}

Oral potentially malignant disorders (OPMDs) represent a critical precursor stage in the development of oral cancer, yet their clinical detection remains challenging due to substantial phenotypic heterogeneity and overlap with benign conditions. While image-based deep learning approaches show promise for automated screening, visual information alone is often insufficient in real-world clinical settings where diagnostic decisions rely on both lesion appearance and patient-specific risk factors. We propose a multimodal deep learning framework for OPMDs (M2-OPMDNet), that jointly integrates co-registered white-light and autofluorescence intraoral images with systematically curated clinical information for OPMD detection. A customized, structured questionnaire was designed to capture clinically meaningful risk factors and symptomatology in a standardized and reproducible manner, enabling seamless integration with image-derived features. To assess the impact of representation learning strategies, we evaluated multiple image encoder backbones, including conventional convolutional neural networks and foundation model–based encoders. Model performance was assessed using a prospectively collected dataset reflecting real-world screening conditions. Interpretability was achieved using SHapley Additive exPlanations (SHAP) to quantify feature- and modality-level contributions to model predictions. M2-OPMDNet achieved an AUC of 0.952, outperforming unimodal approaches and demonstrating improved robustness for visually subtle lesions. SHAP analysis showed that structured clinical variables contributed substantially to risk estimation and complemented image-based features. This study demonstrates that explainable multimodal learning combining white-light and autofluorescence imaging with structured clinical data enables accurate, transparent, and clinically grounded detection of OPMDs. By unifying optimized data acquisition, standardized clinical questionnaires, and interpretable multimodal deep learning, M2-OPMDNet provides a scalable and trustworthy framework for real-world oral cancer screening and decision support.
\end{abstract}
\end{frontmatter}
\section{Introduction}
Oral cancer remains a major global health burden, with an estimated 389,485 new cases and 188,230 deaths worldwide in 2022 (\cite{bray2024global}). Its incidence has steadily increased over recent decades, underscoring the urgent need for earlier and more reliable detection strategies (\cite{conway2018changing,stepan2023changing,miranda2020global}). A substantial proportion of oral cancers develop from oral potentially malignant disorders (OPMDs), a heterogeneous group of clinically identifiable lesions that carry an elevated risk of malignant transformation. OPMDs exhibit wide variability in color, surface texture, morphology, size, and anatomical location, and several subtypes closely resemble benign or inflammatory conditions (\cite{warnakulasuriya2020oral,speight2018oral}). This phenotypic diversity complicates accurate risk stratification in routine clinical practice and frequently contributes to delayed diagnosis. Effective screening and timely identification of OPMDs are therefore critical to reducing progression to invasive cancer and improving patient survival (\cite{warnakulasuriya2021oral,thankappan2021cost}).

Photographic imaging has emerged as a practical, non-invasive modality for lesion documentation and preliminary triage, owing to its ability to capture salient visual characteristics of oral mucosal abnormalities (\cite{you2025real,wong2019using}). In addition to conventional white-light imaging, tissue autofluorescence imaging has gained attention as a complementary technique for oral cancer and OPMD screening. Autofluorescence exploits endogenous fluorophores within oral tissues, such as collagen and flavins, whose emission patterns are altered by dysplasia-related changes in epithelial thickness, metabolic activity, and stromal integrity (\cite{bhokare2025diagnostic,shi2019potential,bodhade2026efficacy}). Under specific excitation wavelengths, potentially malignant or malignant lesions often exhibit loss of fluorescence relative to surrounding healthy mucosa. This contrast enhancement can reveal subclinical or visually subtle abnormalities that may be inconspicuous under white light alone. As a result, combined white-light and autofluorescence imaging offers a richer representation of lesion characteristics and has shown promise in improving early detection sensitivity (\cite{uthoff2018point,wang2022diagnostic}). Beyond visible-range photography and autofluorescence, label-free deep-ultraviolet (DUV) imaging (\cite{chen2025label}) has also shown strong potential for enhancing epithelial and stromal contrast in tissue, including improved visualization of nuclei and fibrous structures. Recent developments in multimodality DUV tissue imaging and polarization-resolved DUV microscopy further suggest that additional intrinsic contrast channels may improve sensitivity to subtle precancer-related changes (\cite{you2025self,chen2026polarization}). 

Recent advances in deep learning have further accelerated the development of automated image-based screening systems for oral cancer and OPMDs (\cite{welikala2020automated,warin2021automatic,song2021mobile}). Convolutional neural networks (CNNs), have demonstrated strong capability in capturing local texture and color hues (\cite{desai2021anatomization,zuluaga2021cnn,song2021mobile}), while large-scale foundation models, including CLIP-based architectures, offer powerful generalization through pretraining on massive natural image–text corpora (\cite{pai2024foundation,song2025integrating}). Despite their success in generic vision tasks, the suitability of these foundation models for fine-grained, domain-specific medical imaging—particularly for subtle and heterogeneous lesions such as OPMDs—has not been systematically investigated. Moreover, diagnostically meaningful cues in intraoral images are often spatially localized, modality-dependent, and sensitive to acquisition variability, posing additional challenges to representation learning.

Image-based approaches, although promising, can be compromised by variations in illumination, image quality, lesion size, and acquisition devices(\cite{maron2021benchmark,jaspers2024robustness,you2025real}). Importantly, clinical diagnosis in real-world settings rarely relies on visual appearance alone. Clinicians routinely integrate imaging findings with patient-reported symptoms and risk factors, such as tobacco or alcohol use, lesion duration, pain, ulceration, and prior medical history(\cite{kar2020improvement,jain2024oral,mavedatnia2023oral}). Parallel to image-based research, several studies have highlighted the predictive value of demographic and clinical variables, including age, sex, lifestyle habits, lesion site, and symptomatology. While such tabular data provide valuable contextual information, models based solely on clinical variables often lack the specificity needed for accurate lesion-level assessment. Recent attempts to combine oral images with selected clinical features suggest that multi-modal learning can outperform unimodal approaches (\cite{devindi2024multimodal,song2018automatic}).

In this work, we jointly model multi-modal imaging data and systematically curated clinical information. We introduce a multi-modal neural network for OPMD detection (M2-OPMDNet) together with an intraoral imaging device capable of acquiring co-registered white-light and autofluorescence images (\cite{uthoff2019small,birur2022field}). To ensure consistent and clinically meaningful data collection, we designed and implemented a customized patient questionnaire tailored specifically for OPMD screening. Unlike free-text clinical narratives, which are often heterogeneous, incomplete, and difficult to standardize, the structured questionnaire enables systematic capture of domain-specific risk factors and symptom patterns using predefined and clinically interpretable items. This design not only reduces ambiguity and missing information but also facilitates reliable quantitative modeling and seamless integration with image-derived features. By embedding knowledge of OPMD-related information into the questionnaire structure, our approach provides a principled and reproducible mechanism for encoding clinical context that is directly optimized for multi-modal learning, rather than relying on generic language-based representations or unstructured clinical notes. The standardized design of this questionnaire enables effective integration of structured clinical data with image-derived features, facilitating robust multi-modal learning.

Furthermore, to elucidate how architectural choices influence diagnostic performance, we systematically evaluated M2-OPMDNet using a diverse set of image encoder backbones, spanning conventional CNNs (ResNet-50, VGG-19, EfficientNet-B4, and DenseNet-121) and foundation model–based encoders derived from CLIP (CLIP-ResNet-50 and CLIP-ViT)(\cite{radford2021learning}). These backbones embody distinct design philosophies, ranging from deep sequential convolution and parameter-efficient scaling to transformer-based global attention mechanisms. By training each configuration on both bright-field and autofluorescence images, with and without the questionnaire modality, we provide a comprehensive analysis of representation learning strategies for OPMD detection, offering practical insights into balancing diagnostic accuracy, robustness, and computational efficiency for real-world screening deployment.

Using a prospectively collected dataset that reflects real-world screening conditions(\cite{yang2023performance}), M2-OPMDNet achieves an area under the ROC curve (AUC) of 0.952 for early OPMD detection, demonstrating strong discriminative performance. To enhance transparency and clinical relevance, we further apply SHapley Additive exPlanations (SHAP) to provide fine-grained interpretability at both the modality and feature levels (\cite{lundberg2018explainable}). Specifically, SHAP enables quantitative attribution of model predictions to individual questionnaire items and image-derived features, thereby revealing which clinical questions and which data modalities most strongly influence diagnostic decisions (\cite{nohara2022explanation,li2024interpretable,miao2024exploring,laatifi2023explanatory}).

At the questionnaire level, SHAP analysis identifies the relative importance of each structured clinical item, allowing systematic ranking of patient-reported and clinician-recorded factors according to their contribution to OPMD risk estimation. This capability offers a data-driven mechanism to determine which questions carry the greatest predictive value and which provide limited or redundant information, directly informing questionnaire refinement, clinical interview prioritization, and streamlined screening workflows. Notably, one of the most salient findings is that current tobacco use consistently emerges as the highest-impact questionnaire variable, aligning with established epidemiological evidence and reinforcing the biological plausibility of the model’s decision process (\cite{jiang2019tobacco,chaturvedi2019tobacco}). Additional influential factors include lesion persistence, pain, ulceration, and alcohol consumption, collectively capturing key behavioral and symptomatic dimensions of OPMD risk.

Beyond individual clinical variables, SHAP further enables modality-level attribution, disentangling the relative contributions of imaging and questionnaire inputs to overall predictions. This analysis reveals how structured clinical information complements visual cues, particularly in cases where lesion appearance is subtle or ambiguous. By quantifying the dynamic balance between image-based and questionnaire-driven evidence, SHAP provides critical insights into multi-modal fusion behavior, demonstrating that clinical context can substantially compensate for limitations in image-only representations and improve robustness across diverse lesion presentations.
Together, these interpretability results establish a transparent link between model predictions, clinical knowledge, and patient-specific risk factors. This framework not only strengthens trust in automated decision support but also creates a closed-loop mechanism for continuously improving both questionnaire design and model architecture. By highlighting the most informative clinical questions and the relative importance of each modality, SHAP-guided analysis supports the development of more efficient, explainable, and clinically actionable OPMD screening systems.

\begin{figure}[hbtp]
    \centering
    \includegraphics[width=\columnwidth]{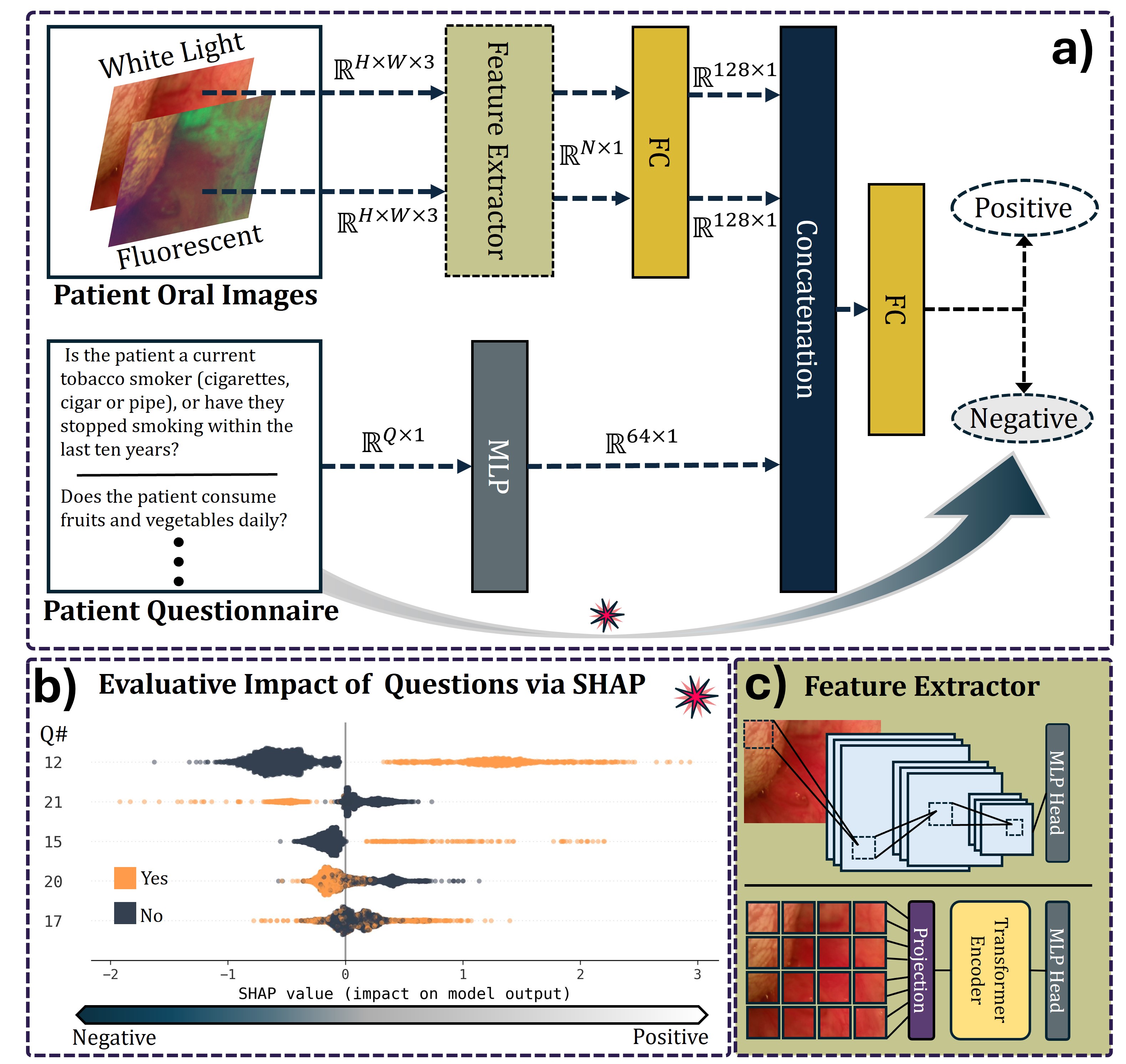}
    \caption{Overview of Multi-Modal Oral Potentially Malignant Disorders (OPMDs) Detection Nerual Network (M2-OPMDNet). a) The multi-modal oral image network captures both white-light and fluorescent images of the exact same intraoral region. These images are passed into feature extractors to generate embeddings, which are then concatenated with questionnaire embeddings for final classification. b) The top 5 questionnaire items with the greatest impact on the final results are shown. The color indicates the patient’s answer: yes (orange) or no (black). The x-axis represents how each question influences the classification decision, either pushing it toward a more positive or negative outcome. Each point corresponds to a patient case. All 23 questions are listed in the appendix. c) We compared classification accuracy across different backbone structures for the feature extractors, specifically between convolutional neural networks and attention-based transformers.}
    \label{fig1:nerual network overview} 
\end{figure}

\begin{figure}[hbtp]
    \centering
    \includegraphics[width=\columnwidth]{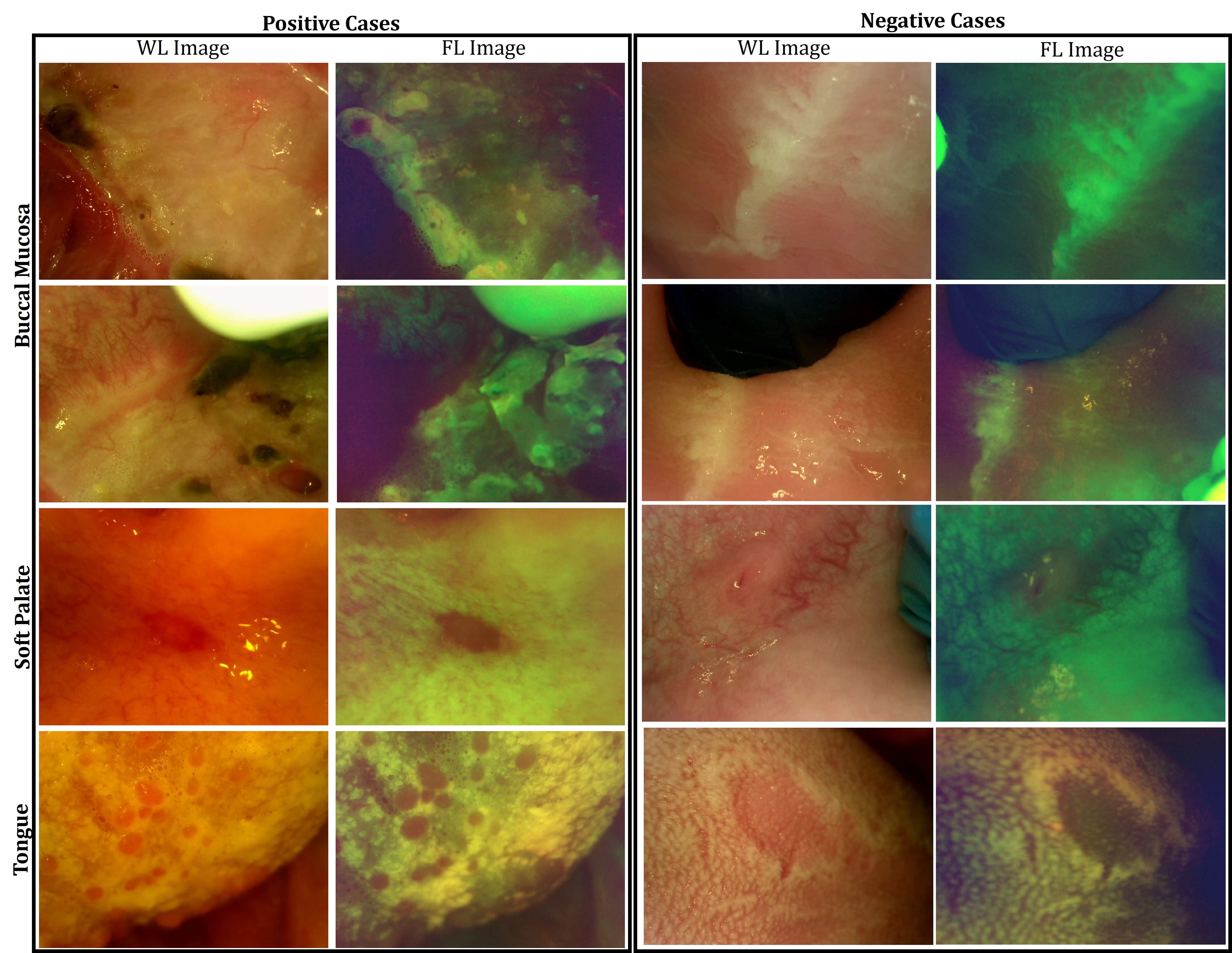}
    \caption{Comparison between white-light and fluorescent images across various oral cavity regions. The left panel shows positive cases, while the right panel shows negative cases.}
    \label{fig2:dataset show} 
\end{figure}

\section{Methods}

\subsection{Multi-Modal OPMD Detection Network}
M2-OPMDNet is a multi-modal neural network designed to integrate intraoral imaging data with structured patient questionnaire information for OPMD detection. The network consists of two parallel feature extraction modules, image encoder and a questionnaire encoder, followed by a fusion and classification module.

Image Encoder: The image encoder processes intraoral images acquired under white-light and autofluorescence illumination. Images are first normalized and resized to a fixed spatial resolution before being fed into a convolutional backbone network. The backbone extracts hierarchical visual features capturing lesion color, texture, and structural patterns relevant to OPMD characterization. The final convolutional feature maps are globally aggregated and projected through a series of fully connected (FC) layers to obtain a compact image representation of dimension \(B \times 128 \times 1\), where \(B\) denotes the batch size. Nonlinear activation functions and dropout are applied between FC layers to enhance representation capacity and mitigate overfitting.

Questionnaire Encoder: The questionnaire encoder is designed to process structured patient information derived from the customized clinical questionnaire. Each questionnaire item is encoded as a binary variable and concatenated into a fixed-length input vector. This vector is passed through a multilayer perceptron (MLP) composed of successive FC layers with nonlinear activations. The MLP learns higher-order interactions among questionnaire items and maps the input into a latent embedding of size \(B \times 64 \times 1\). This design allows the model to capture both individual risk factors and their combinatorial effects while maintaining interpretability at the feature level.

Multi-modal Feature Fusion and Classification: The image and questionnaire embeddings are concatenated along the feature dimensions to form a unified representation of size \(B \times 192 \times 1\). This fused embedding is subsequently passed to a final FC classification head, which consists of two FC layers followed by a sigmoid activation function to produce a scalar probability representing the likelihood of OPMD presence. During training, the network is optimized end-to-end using a cross-entropy loss function.
This modular architecture enables flexible integration of complementary visual and clinical information, while preserving clear feature attribution pathways for downstream interpretability analysis. In particular, the separation of image and questionnaire encoders facilitates SHAP-based assessment of the relative contribution of each modality and individual questionnaire items to the final prediction.

\subsection{SHapley Additive exPlanations}
SHAP (\cite{lundberg2017unified} has gained considerable traction in interpretable machine learning, particularly in sensitive domains such as medicine (\cite{lundberg2018explainable}), where understanding the impact of input variables (e.g., clinical biomarkers, imaging features, questionnaire responses) on a diagnostic or prognostic output is critical. SHAP not only satisfies desirable axioms such as local accuracy, consistency, and missingness but also provides a unified approach to interpret complex models including deep neural networks, gradient-boosted trees, and ensemble models.

Let $f: \mathbb{R}^n \to \mathbb{R}$ be a machine learning model that maps an input feature vector $\mathbf{x} = [x_1, x_2, \ldots, x_n]$ to an output (e.g., a predicted probability or score). The goal of SHAP is to assign each input feature $x_i$ a real-valued attribution $\phi_i$ such that:

\begin{equation}
f(\mathbf{x}) = \phi_0 + \sum_{i=1}^{n} \phi_i
\end{equation}

\noindent where $\phi_0$ is the expected model output over the background dataset:

\begin{equation}
\phi_0 = \mathbb{E}_{\mathbf{x}'}[f(\mathbf{x}')]
\end{equation}

\noindent The SHAP value $\phi_i$ for feature $i$ is defined as the average marginal contribution of $x_i$ to all possible subsets $S \subseteq \{1, \ldots, n\} \setminus \{i\}$ of the feature set:

\begin{equation}
\phi_i = \sum_{S \subseteq N \setminus \{i\}} \frac{|S|! (n - |S| - 1)!}{n!} \left[ f_{S \cup \{i\}}(\mathbf{x}) - f_S(\mathbf{x}) \right]
\end{equation}

\noindent where the $N$ is the set of all feature indices $\{1, 2, \ldots, n\}$, $f_S(\mathbf{x})$ denotes the model prediction when only the subset of features $S$ is known (and the rest are marginalized over a background distribution), and the weighting term $\frac{|S|! (n - |S| - 1)!}{n!}$ ensures fairness by considering all possible feature orderings.

In medical settings, SHAP has gained attention of applied to interpret black-box models in tasks such as disease classification, treatment outcome prediction, and clinical decision support. By assigning interpretable scores to each input (e.g. patient age, imaging biomarker intensity, or genetic profile), SHAP can facilitate both trust and accountability to assist diagnostic results of the deep learning model, which are essential for clinical deployment. In our M2-OPMD Net setting, SHAP allows for both individual cases and global explanations, enabling clinicians to understand model behavior at both the individual and population level. This is particularly useful when combined with structured tabular inputs and multimodal data, as it greatly improves interpretation of model outputs for each patient, making the results more reliable and interpretable for precise OPMD diagnosis.

\subsection{SHAP-Based Interpretability for the Multi-Modal Prediction}

To interpret the predictions of the proposed multi-modal network and quantify the contribution of individual input features, we employed SHapley Additive exPlanations (SHAP), a model-agnostic framework grounded in cooperative game theory. SHAP attributes a prediction to individual input features by computing Shapley values, which represent each feature’s marginal contribution to the model output relative to a reference baseline.
Application to Multi-Modal Architecture: Given the dual-branch architecture of M2-OPMDNet, SHAP analysis was conducted jointly on both the image and questionnaire modalities while preserving their respective input structures. The model output—defined as the predicted probability of OPMD presence—served as the target function for explanation. For each test sample, SHAP values were computed to estimate the contribution of (i) image-derived latent features from the image encoder and (ii) individual binary questionnaire items processed by the questionnaire encoder.
Questionnaire-Level Attribution: For the questionnaire modality, SHAP values were calculated at the level of individual questionnaire items, enabling direct attribution of the prediction to specific patient-reported or clinician-recorded responses. Each input was treated as independent features, and a background distribution was constructed from a representative subset of the training data to approximate baseline clinical profiles. This approach allows identification of the questionnaire questions most strongly associated with elevated or reduced OPMD risk, facilitating both population-level and case-level interpretability.
Multi-Modal Contribution Analysis: By aggregating SHAP values across samples, we assessed the relative importance of the modalities and examined how visual and clinical features jointly influence predictions. SHAP summaries were used to rank questionnaire items by importance and to compare their contribution against image-derived features. 
This SHAP-based interpretability framework supports transparent deployment of the proposed system by enabling clinicians to understand not only what the model predicts, but why. Importantly, identifying the most influential questionnaire items provides a data-driven mechanism to refine patient questionnaires, prioritize clinically meaningful questions, and inform future model and data collection design. All SHAP analyses were performed post hoc on the trained model and did not affect model optimization.

\subsection{Data Collection}
A total of 2650 paired white light and autofluoresence images of healthy oral mucosa and oral lesions suspicious for OPML or OC were collected from 667 subjects, together with a full documentation of any clinical signs and symptoms as well as an evaluation of all individual risk factors and behaviors. Images and data were recorded using the customized intraoral camera system and software specifically designed for non-specialists working in a low resource community setting. This study was performed in 667 subjects from three clinical sites. Individuals attending Concorde College of Dental Hygiene Dental Clinics in Garden Grove, California, or West Coast University Dental Hygiene Clinics in Anaheim, California, as well as those referred to the University of California, Irvine Clinics with oral lesions suspicious of OPML or OC were recruited, as well as individuals with no visible oral lesions. Informed consent was obtained from all subjects prior to study begin. Examples of the white light and auto-fluoresence images in this dataset were shown in Fig. \ref{fig2:dataset show}. The OPMD/OC risk and symptom questionnaire list in this dataset was shown in Table~\ref{tab:oral-risk}. 

Dental Hygiene students at Concorde College of Dental Hygiene Dental Clinics in Garden Grove, California, and West Coast University Dental Hygiene Clinics in Anaheim, California and Biomedical Science students at the University of California, Irvine, recorded multimodality image sets, clinical signs and symptoms, and risk factor data from each subject using the screening platform. These students were selected to represent community health workers, who are the intended future users of the screening platform. At each study site, the students all received one day of classroom teaching by the same instructor, where they learned about OPML and OC causes, risk factors, and pathology, as well as its clinical manifestations, prognosis, treatment, and outcomes. Next, directly prior to study begin, the students all attended a half day of clinical training at their respective school, which was taught by the same instructor at all centers. During this clinic, the students learned to operate the screening platform’s prototype scanner pen and App. Throughout the study duration, the students recorded study participants multimodality images, risk factors, clinical signs and symptoms with the scanner pen and App. Imaging sites were selected in each subject as follows: in all subjects with lesions, all soft tissue areas manifesting any kind of visual changes were imaged. Then, where available, contralateral lesion-free tissues were also imaged in the same subject. In healthy subjects, 12 intraoral sites were imaged, including buccal and vestibular mucosa, dorsal, ventral, and lateral surfaces of tongue, floor of mouth, palate, soft palate, and tonsillar regions. Finally, one oral medicine specialist performed a full standard of care OC screening and subsequently recorded a screening outcome as either “no increased risk” or “increased risk” for each study participant. During this project the same oral medicine specialist performed all screening in all participants at all three clinical sites according to the standard of care, combining clinical examination with risk factors and patient history. The specialist screening outcome served as the gold standard for evaluating screening platform accuracy. Finally, all study participants with increased OC risk according to the specialist screening were informed of the screening outcome. They were then referred to a specialist for diagnosis and entry into the pathway of care.

\section{Experiment}

\subsection{Training}
We trained M2-OPMDNet on an in-house dataset of co-registered white-light and autofluorescence images captured from the exact same field of view (FOV). This co-registration ensures feature correspondence across modalities, facilitating reliable OPMD detection. The text branch ingests a clinician-administered 23-item questionnaire completed by each patient.

Each patient contributed images from multiple intraoral regions; some regions were OPMD-positive and others negative. To avoid label mismatch between region-level images and patient-level responses, we paired questionnaires only with images from the same patient whose regional label was concordant with the clinical finding. To prevent information leakage, cross-validation splits were created at the patient level rather than the image level. Folds were stratified to balance the number of image pairs across splits, though the number of patients per fold may vary.
To isolate feature extractor architecture impact to the final performance, all models were trained under identical hyperparameters and procedures. Image backbones were initialized from \textit{ImageNet-21K} pretraining (CLIP used its released contrastive pretraining). Optimization employed Adam ($\beta_{1}=0.9$, $\beta_{2}=0.99$, $\epsilon=10^{-8}$) with weight decay $0.1$. We used a two-stage schedule: (i) head-only training for 20 epochs at a learning rate of $1\times10^{-5}$ with all image backbones frozen, followed by (ii) full fine-tuning with the backbones unfrozen at a reduced learning rate of $1\times10^{-6}$. The training loss function was class-weighted binary cross-entropy. To preserve cross-modality image spatial features, we applied synchronized augmentations to each white-light/fluorescence image pair—identical random crop, rotation, and color-jitter parameters were used for both images—while no augmentation was applied to the questionnaire/text branch.

\subsection{Encoder comparsion}
\begin{figure}[hbtp]
    \centering
    \includegraphics[width=\columnwidth]{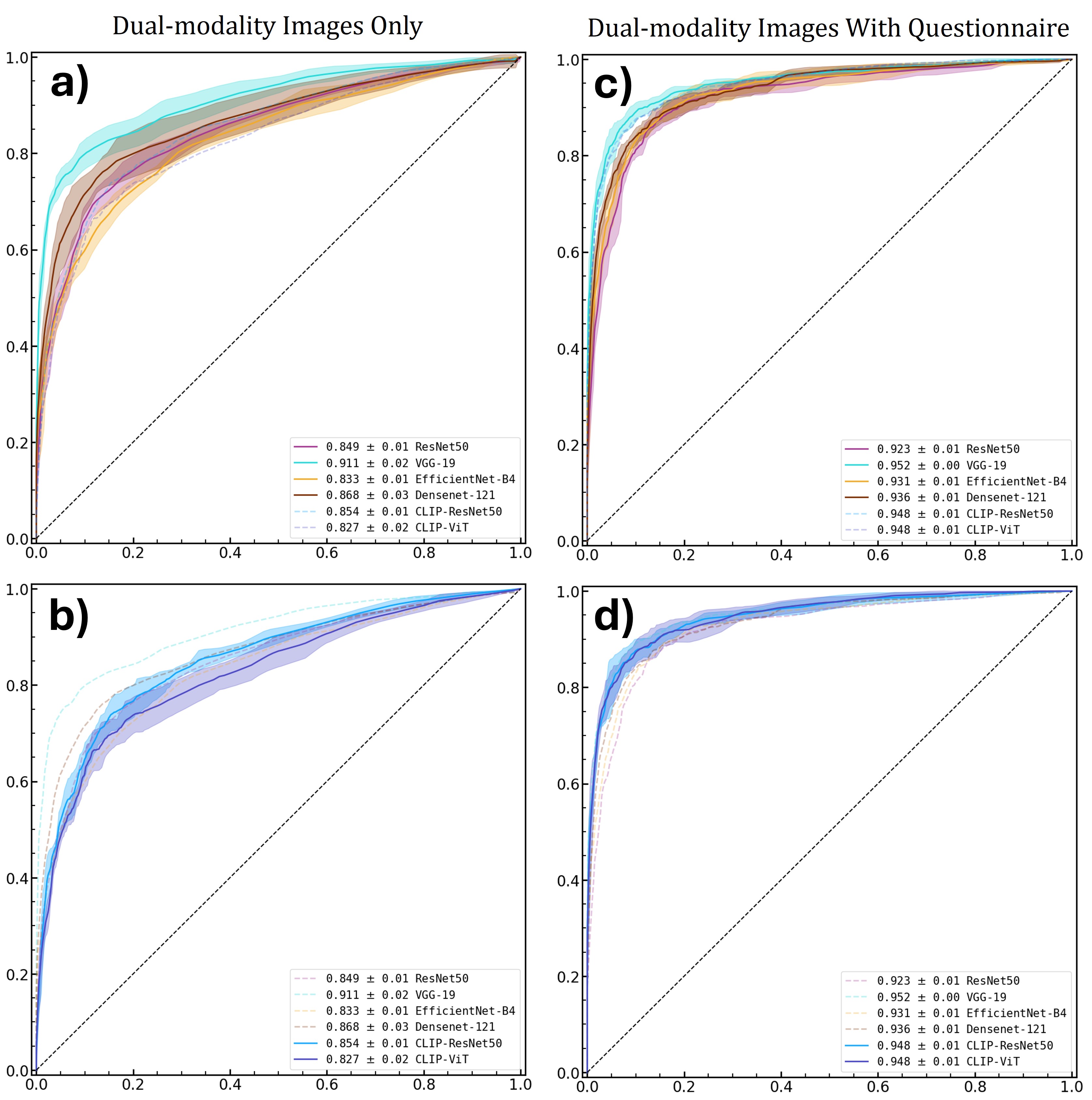}
    \caption{ROC curve comparison of M2-OPMDNet with and without questionnaire input. All results are obtained using five-fold cross-validation, with the mean shown as a solid line and the variance indicated by the shaded error region. The area under the curve (AUC) for each image-extractor backbone is reported in the legend. (a) and (b) present results using only dual-modality oral images as input. (c) and (d) show results from the multi-modal neural network that incorporates both dual-modality oral images and questionnaire data.}
    \label{fig3:roc_curve compare} 
\end{figure}

We systematically evaluated M2-OPMDNet using a diverse set of image encoder backbones to examine how architectural choices influence OPMD detection performance. Specifically, we compared conventional convolutional neural networks (CNNs), including ResNet-50 (\cite{he2016deep}), VGG-19 (\cite{simonyan2014very}), EfficientNet-B4 (\cite{tan2019efficientnet}), and DenseNet-121 (\cite{huang2017densely}), with large-scale foundation model–based encoders derived from CLIP, CLIP-ResNet-50 (\cite{radford2021learning}) and CLIP-ViT. These backbones represent distinct design philosophies in visual representation learning, ranging from deep sequential convolutional architectures to parameter-efficient scaling strategies and transformer-based global attention mechanisms.
To assess the robustness of each backbone, M2-OPMDNet was trained using each image encoder to extract features from both bright-field and auto-fluorescence intraoral images. This evaluation is particularly relevant in the context of OPMD detection, where diagnostically meaningful cues may be subtle, spatially localized, and modality-dependent. The comparative analysis also provides practical insights into selecting image encoders that achieve reliable diagnostic accuracy while balancing computational complexity and resource constraints, which are critical considerations for deployment in real-world screening environments.
Diagnostic performance was quantified using the area under the receiver operating characteristic curve (AUC), as summarized in Fig.~\ref{fig3:roc_curve compare}. We further examined the contribution of multi-modal fusion by training and validating each model configuration with and without the questionnaire modality. In the image-only setting, VGG-19 consistently achieved the highest AUC among the evaluated backbones, outperforming ResNet-50, EfficientNet-B4, and DenseNet-121. This result suggests that the VGG architecture’s deep yet relatively simple convolutional structure is well suited for capturing fine-grained texture and color variations in intraoral images, including fluorescence intensity patterns associated with OPMDs.
In contrast, the CLIP-based foundation models—both CLIP-ResNet-50 and CLIP-ViT—demonstrated lower performance when operating on image data alone, trailing VGG-19 by approximately $8\%$ AUC. This observation indicates that, despite their strong generalization capabilities on large-scale natural image datasets, foundation models may be less effective for specialized medical imaging tasks when domain-specific supervision is limited. Notably, the inclusion of the questionnaire modality substantially improved performance across all architectures, with the most pronounced gains observed for the CLIP-based encoders. When clinical questionnaire data were incorporated, CLIP-based models perform much better than image-only training.

\subsection{SHAPE-based assessment of M2-OPMDNet}

M2-OPMDNet was interpreted at two complementary levels. First, we performed case-level attribution analysis to explain individual predictions by quantifying the relative contributions of the two imaging modalities and the questionnaire inputs for each patient. Second, we conducted a dataset-level analysis focused on the questionnaire features, aggregating SHAP attributions across the entire cohort using 5-fold cross-validation to rank questionnaire items by their overall influence on the model output. Together, these two views provide both localized, patient-specific explanations and a global assessment of which questionnaire factors most strongly drive M2-OPMDNet’s diagnostic decisions, improving transparency and supporting clinically meaningful interpretation.

\begin{figure}[hbtp]
    \centering
    \includegraphics[width=\columnwidth,height=0.78\textheight,keepaspectratio]{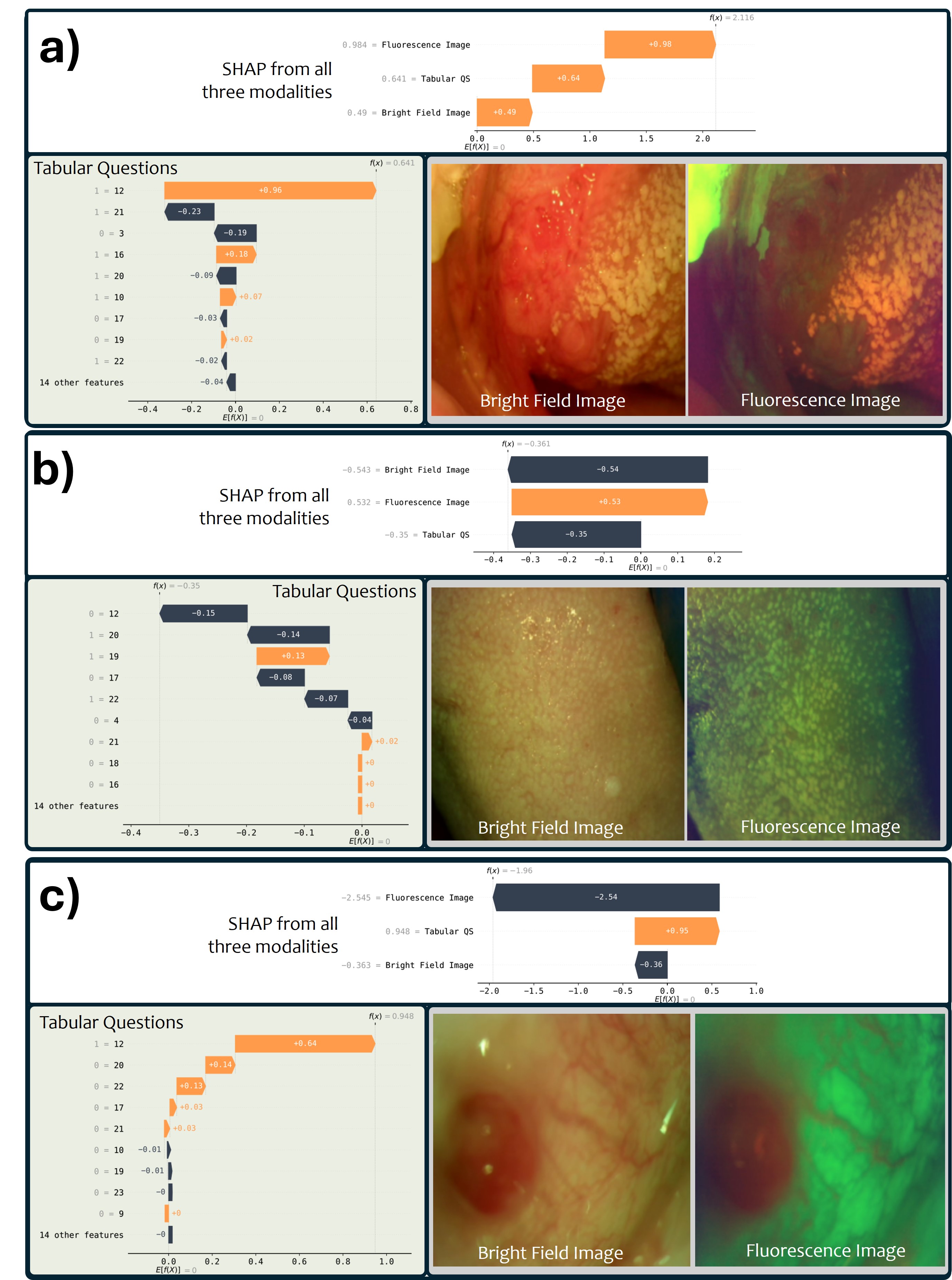}
    \caption{SHAP case studies on the impact of all modalities. Panels (a–c) show three individual cases from a trained M2-OPMDNet using VGG-19 as the image feature extractor, illustrating how each modality influences the final prediction. These explanations provide clinicians with additional insight before making a final OPMD diagnosis when using a neural-network classifier.The full list of the questions is provided in the appendix. Table~\ref{tab:oral-risk}}
    \label{fig4:case_stuides} 
\end{figure}

\begin{figure}[hbtp]
    \centering
    \includegraphics[width=\columnwidth]{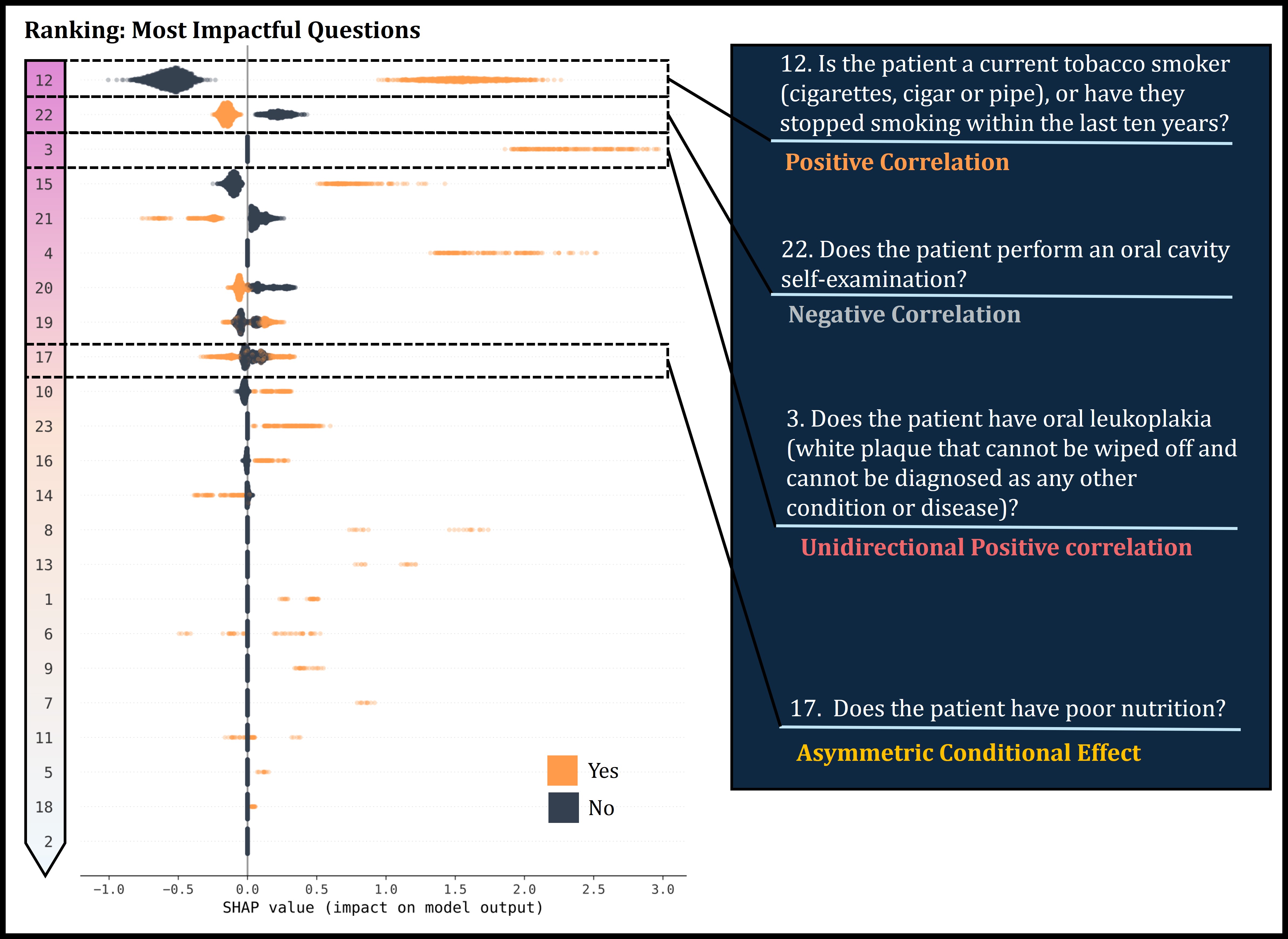}
    \caption{Ranking of the most impactful clinical questions (n=23) on neural network outputs via SHAP. We group each question’s effect into four relationships: positive correlation, negative correlation, unidirectional positive correlation, and asymmetric conditional effect. The full list of OPMD-related clinical questions is provided in the appendix Table~\ref{tab:oral-risk}}
    \label{fig5:qs_ranking} 
\end{figure}

For individual cases, we present SHAP waterfall plots that illustrate how image features and questionnaire variables push the prediction toward OPMD or non-OPMD, providing clinician-readable explanations. We find SHAP to be a useful tool for guiding clinical decisions by revealing the main sources of influence from the input variables. Clinicians can trace which modalities and features contributed to each classification, enabling a review of both their own assessment and the model’s output for more accurate, rigorous decisions in cases where OPMD can be confused with non-precancer symptoms. We show three case studies in Fig.~\ref{fig4:case_stuides}: all are true-positive predictions by M2-OPMDNet, illustrating why the network made each decision and how the questionnaire responses, bright-field images, and fluorescence images contributed to the final result. Case showed in Fig.~\ref{fig4:case_stuides} a) showed unanimous agreement across all three modalities, and we can see how each was weighted toward the final result, indicating that the OPMD signal is most apparent in the fluorescence image. An interesting case in Fig.~\ref{fig4:case_stuides} shows the questionnaire acting as the tie-breaker in the final decision, even though its SHAP contribution is the smallest among the three modalities compared with the bright-field and fluorescence images. Importantly, while the overall prediction is correct, the SHAP attributions reveal that the network interprets the bright-field image as evidence against OPMD, whereas the fluorescence image pushes toward OPMD; this transparency may encourage clinicians to review additional patient images to confirm the diagnosis. In contrast to the previous case, the example in Fig.~\ref{fig4:case_stuides}(c) shows that the questionnaire could have biased the prediction in the wrong direction; however, the bright-field and fluorescence images provided clear evidence that corrected the result. Taken together, the three cases demonstrate how SHAP offers clinician-readable explanations that go beyond a bare “positive” or “negative” label. Additionally, We also aggregate SHAP values across all cases to identify which clinical questions most strongly influence predictions at the population level in Fig.~\ref{fig5:qs_ranking}. We can identify which questions most influence the classifier’s diagnostic results and whether “yes” or “no” responses positively or negatively affect detection. This is extremely useful for guiding clinical decisions, especially for marginal cases between non-OPMD and OPMD. For example, when intraoral symptoms are difficult to determine but the questionnaire suggests a high likelihood of OPMD, clinicians can conduct further examinations and a more careful review to provide a more accurate diagnosis. The chart also quantifies the impact of these tabular questions across a large cohort, providing statistical evidence of their effects. It highlights globally important factors while complementing case-level analyses.

\section{Discussion}
In this study, we proposed M2-OPMDNet, a multi-modal deep learning framework that integrates bright-field intraoral images, autofluorescence images, and structured patient questionnaire data for early detection of oral potentially malignant disorders. Our results demonstrate that combining complementary imaging modalities with targeted clinical information substantially improves diagnostic performance and robustness compared with image-only approaches.
OPMD detection presents unique challenges due to the heterogeneity of lesion appearance and the frequent overlap between premalignant and benign oral conditions. While intraoral photography captures important visual cues, subtle lesions may be difficult to distinguish under white light alone. Autofluorescence imaging enhances contrast by highlighting metabolic and structural tissue alterations, providing additional discriminatory information. Our results show that jointly leveraging bright-field and autofluorescence images yields more reliable representations than either modality alone, particularly when integrated within a unified learning framework.
Importantly, we demonstrate that structured clinical information captured through a customized patient questionnaire plays a critical complementary role. Questionnaire features such as lesion duration, symptoms, and risk behaviors encode contextual knowledge routinely used by clinicians but often omitted from automated screening systems. The substantial performance gains observed when incorporating questionnaire data—especially for foundation model backbones—underscore the importance of aligning model inputs with real-world diagnostic workflows rather than relying exclusively on visual data.
Through a systematic comparison of multiple image backbones, including VGG-19, ResNet-50, EfficientNet-B4, DenseNet-121, CLIP-ResNet-50, and CLIP-ViT, we found that conventional CNN architectures, particularly VGG-19, achieved superior performance in the image-only setting. This suggests that deep convolutional models with strong inductive biases toward local texture and color patterns remain well suited for intraoral imaging tasks, where diagnostic cues are often subtle and spatially localized.
In contrast, transformer-based and foundation models pretrained on large-scale natural image datasets underperformed when used without clinical context. However, once questionnaire data were incorporated, these models closed the performance gap with VGG-based networks. This finding highlights two key points: first, foundation models may require additional structured context to adapt effectively to specialized medical imaging domains; second, multi-modal fusion can mitigate limitations of individual representation paradigms, enabling more flexible backbone selection under varying computational and deployment constraints.
A major strength of this work lies in the use of SHAP to provide transparent, clinician-readable explanations at both the case and population levels. At the individual case level, SHAP waterfall plots reveal how each modality contributes to a specific prediction, enabling clinicians to understand not only the final classification but also the underlying rationale. This transparency is particularly valuable in borderline cases where different modalities provide conflicting signals, encouraging careful review rather than blind trust in automated outputs.
At the population level, aggregating SHAP values across patients identifies which questionnaire items most strongly influence predictions and clarifies whether specific responses increase or decrease estimated risk. This analysis provides empirical support for the clinical relevance of selected questionnaire items and offers a data-driven mechanism to refine and prioritize future questionnaire design. Beyond model interpretability, this feedback loop has the potential to improve patient interviews, focus clinical attention on high-impact risk factors, and inform the design of more efficient and targeted screening protocols.
Several limitations should be acknowledged. First, although the dataset was prospectively collected, its size and demographic composition may limit generalizability across populations and healthcare settings. Second, questionnaire responses were restricted to binary inputs; incorporating ordinal or continuous variables may further enhance risk modeling. Third, while SHAP provides valuable post hoc explanations, it does not guarantee causal interpretation, and its reliability depends on the choice of background distribution and model stability.
Future work will focus on expanding the dataset across multiple centers, incorporating longitudinal follow-up data to assess malignant transformation risk, and exploring adaptive questionnaire designs informed by SHAP-derived importance rankings. On the hardware side, advances in micro-fabrication and additive manufacturing of micro-optics (\cite{hong2021three,hong2022high}) (e.g., multi-photon polymerization and 3D-printed diffractive elements) may enable more compact, low-cost, and scalable oral imaging probes (\cite{you2025extremely,hong2025dual}). Additionally, extending the framework to support real-time deployment and clinician-in-the-loop learning represents a promising direction for translational impact.

\section{Conclusion}
In this study, we present M2-OPMDNet, a clinically grounded multi-modal deep learning framework for OPMD detection that integrates bright-field and autofluorescence intraoral images with structured patient questionnaire data. Our results demonstrate that multi-modal fusion significantly enhances diagnostic performance and robustness compared with image-only approaches, particularly for transformer-based and foundation models. By incorporating SHAP-based interpretability, the proposed system provides transparent, clinician-readable explanations that support both individual case review and population-level insights. Together, these findings highlight the value of combining complementary imaging modalities, targeted clinical information, and explainable AI to enable accurate, trustworthy, and clinically actionable OPMD screening systems.

\appendix
\section{Supporting Information} \label{apx:ap1}
{\rowcolors{2}{red!33!green!46!blue!51}{gray!20}
\setlength{\tabcolsep}{6pt}
\begin{table}[htbp]
\centering
\caption{Oral Cancer/OPMD Risk and Symptom Questionnaire}
\label{tab:oral-risk}
\begin{tabular}{@{}>{\RaggedRight\arraybackslash}p{0.06\textwidth} >{\RaggedRight\arraybackslash}p{0.88\textwidth}@{}}
\toprule
\textbf{\#} & \textbf{Question} \\
\midrule
1  & Has the patient had a previous history of oral squamous cell carcinoma, severe dysplasia/carcinoma-in-situ, or mild/moderate oral dysplasia? \\
2  & Has the patient had previous head/neck radiation therapy for a non-squamous cell carcinoma tumor? \\
3  & Does the patient have oral leukoplakia (white plaque that cannot be wiped off and cannot be diagnosed as any other condition or disease)? \\
4  & Does the patient have oral erythroplakia (red plaque that cannot be diagnosed as any other condition or disease)? \\
5  & Does the patient have a non-healing oral ulcer (greater than two weeks)? \\
6  & Does the patient have any unexplained oral swelling, oral numbness, or tingling sensation? \\
7  & Does the patient have difficulty chewing or moving the jaw or tongue? \\
8  & Does the patient have an unexplained sore throat or feeling that something is caught in the throat, or chronic hoarseness? \\
9  & Does the patient have enlarged (greater than one centimeter) non-painful neck lymph node(s)? \\
10 & Does the patient have lip lesions including white areas, non-healing ulcers or loss of vermillion border integrity? \\
11 & Does the patient have current/history of clinical lichen planus (non-biopsy proven)? \\
12 & Is the patient a current tobacco smoker (cigarettes, cigar or pipe), or have they stopped smoking within the last ten years? \\
\bottomrule
\end{tabular}
\end{table}

\begin{table}[htbp]
\centering
\addtocounter{table}{-1}
\caption{Oral Cancer/OPMD Risk and Symptom Questionnaire (continued)}
\begin{tabular}{@{}>{\RaggedRight\arraybackslash}p{0.06\textwidth} >{\RaggedRight\arraybackslash}p{0.88\textwidth}@{}}
\toprule
\textbf{\#} & \textbf{Question} \\
\midrule
13 & Does the patient use smokeless tobacco (e.g., snus, moist snuff or chewing tobacco)? \\
14 & Is the patient a moderate to heavy alcohol drinker (more than two standard drinks per day)? \\
15 & Does the patient have any other habits such as smoking bidis, kreteks (clove cigarettes), marijuana, use of betel quid (paan) or areca nut? \\
16 & Does the patient experience frequent sunburns of the lip? \\
17 & Does the patient have poor nutrition? \\
18 & Does the patient have oral human papillomavirus colonization (high-risk subtypes 16/18 as confirmed by salivary diagnostics)? \\
19 & Is the patient 50 years of age or older? \\
20 & Does the patient have a history of continuous, professionally-delivered dental care (minimum of once yearly)? \\
21 & Does the patient perform an oral cavity self-examination? \\
22 & Does the patient consume fruits and vegetables daily? \\
23 & Is the patient immunocompromised in any way either from medications or a condition? \\
\bottomrule
\end{tabular}
\end{table}}

\pagebreak

\medskip \noindent
\textbf{} \par 

\medskip \noindent
\textbf{Acknowledgements} \par 

\medskip \noindent
\textbf{} \par
This work was supported by NIH (R01DE030682 National Institute of Dental and Craniofacial Research, R21CA274717 National Cancer Institute, U01CA279862 National Cancer Institute).

\medskip \noindent
\textbf{Data Availability Statement} \par 
The data that has been used in this research will not be made public. The code can be obtained upon request.

\medskip \noindent
\textbf{Declaration of competing interest} \par 
The authors declare that they have no known competing financial interests or personal relationships that could have appeared to influence the work reported in this paper.
\medskip

%
\bibliographystyle{elsarticle-harv} 
\bibliography{MMOralNet}

@article{lundberg2018explainable,
  title={Explainable machine-learning predictions for the prevention of hypoxaemia during surgery},
  author={Lundberg, Scott M and Nair, Bala and Vavilala, Monica S and Horibe, Mayumi and Eisses, Michael J and Adams, Trevor and Liston, David E and Low, Daniel King-Wai and Newman, Shu-Fang and Kim, Jerry and others},
  journal={Nature biomedical engineering},
  volume={2},
  number={10},
  pages={749--760},
  year={2018},
  publisher={Nature Publishing Group UK London}
}

@article{you2025self,
  title={Self-calibrating Fourier ptychographic microscopy using automatic differentiation},
  author={You, Ruilin and Liang, Rongguang},
  journal={Optics Letters},
  volume={50},
  number={2},
  pages={415--418},
  year={2025},
  publisher={Optica Publishing Group}
}

@article{chen2026polarization,
  title={Polarization resolved deep ultraviolet microscopy for label free imaging with enhanced nuclei and fiber contrast},
  author={Chen, Jiabin and You, Ruilin and Contreras, Marco and Cai, Haijiang and Sun, Yuanyuan and Wang, Yihan and Song, Bofan and Pau, Stanley and Hong, Zhihan and Liang, Rongguang},
  journal={Optics and Lasers in Engineering},
  volume={196},
  pages={109375},
  year={2026},
  publisher={Elsevier}
}

@article{chen2025label,
  title={Label-free surface sectioning deep ultraviolet tissue imaging in multimodalities},
  author={Chen, Jiabin and You, Ruilin and Contreras, Marco and Cai, Haijiang and Villarreal, Paula Patricia and Vargas, Gracie and Hong, Zhihan and Liang, Rongguang},
  journal={Biomedical Optics Express},
  volume={16},
  number={7},
  pages={2756--2766},
  year={2025},
  publisher={Optica Publishing Group}
}

@article{lundberg2017unified,
  title={A unified approach to interpreting model predictions},
  author={Lundberg, Scott M and Lee, Su-In},
  journal={Advances in neural information processing systems},
  volume={30},
  year={2017}
}

@article{chaturvedi2019tobacco,
  title={Tobacco related oral cancer},
  author={Chaturvedi, Pankaj and Singh, Arjun and Chien, Chih-Yen and Warnakulasuriya, Saman},
  journal={Bmj},
  volume={365},
  year={2019},
  publisher={British Medical Journal Publishing Group}
}

@article{jiang2019tobacco,
  title={Tobacco and oral squamous cell carcinoma: a review of carcinogenic pathways.},
  author={Jiang, Xiaoge and Wu, Jiaxin and Wang, Jiexue and Huang, Ruijie},
  journal={Tobacco Induced Diseases},
  volume={17},
  pages={29},
  year={2019},
  doi={10.18332/tid/105844}
}

@article{laatifi2023explanatory,
  title={Explanatory predictive model for COVID-19 severity risk employing machine learning, shapley addition, and LIME},
  author={Laatifi, Mariam and Douzi, Samira and Ezzine, Hind and Asry, Chadia El and Naya, Abdellah and Bouklouze, Abdelaziz and Zaid, Younes and Naciri, Mariam},
  journal={Scientific Reports},
  volume={13},
  number={1},
  pages={5481},
  year={2023},
  publisher={Nature Publishing Group UK London}
}

@article{miao2024exploring,
  title={Exploring explainable machine learning and Shapley additive exPlanations (SHAP) technique to uncover key factors of HNSC cancer: an analysis of the best practices},
  author={Miao, Kexin and Hounye, Alphonse Houssou and Su, Liuyan and Pan, Qi and Wang, Jiaoju and Hou, Muzhou and Xiong, Li},
  journal={Biomedical Signal Processing and Control},
  volume={89},
  pages={105752},
  year={2024},
  publisher={Elsevier}
}

@article{li2024interpretable,
  title={Interpretable mortality prediction model for ICU patients with pneumonia: using shapley additive explanation method},
  author={Li, Jiaxi and Zhang, Yu and He, ShengYang and Tang, Yan},
  journal={BMC Pulmonary Medicine},
  volume={24},
  number={1},
  pages={447},
  year={2024},
  publisher={Springer}
}

@article{nohara2022explanation,
  title={Explanation of machine learning models using shapley additive explanation and application for real data in hospital},
  author={Nohara, Yasunobu and Matsumoto, Koutarou and Soejima, Hidehisa and Nakashima, Naoki},
  journal={Computer Methods and Programs in Biomedicine},
  volume={214},
  pages={106584},
  year={2022},
  publisher={Elsevier}
}

@article{devindi2024multimodal,
  title={Multimodal deep convolutional neural network pipeline for AI-assisted early detection of oral cancer},
  author={Devindi, GAI and Dissanayake, DMDR and Liyanage, SN and Francis, FBAH and Pavithya, MBD and Piyarathne, NS and Hettiarachchi, PVKS and Rasnayaka, RMSGK and Jayasinghe, Ruwan Duminda and Ragel, Roshan G and others},
  journal={IEEE Access},
  year={2024},
  publisher={IEEE}
}

@article{mavedatnia2023oral,
  title={Oral cancer screening knowledge and practices among dental professionals at the University of Toronto},
  author={Mavedatnia, Dorsa and Cuddy, Karl and Klieb, Hagen and Blanas, Nick and Goodman, Jade and Gilbert, Melanie and Eskander, Antoine},
  journal={BMC Oral Health},
  volume={23},
  number={1},
  pages={343},
  year={2023},
  publisher={Springer}
}

@article{jain2024oral,
  title={Oral cancer screening: insights into epidemiology, risk factors, and screening programs for improved early detection},
  author={Jain, Amit Kumar},
  journal={Cancer Screening and Prevention},
  volume={3},
  number={2},
  pages={97--105},
  year={2024},
  publisher={Xia \& He Publishing Inc.}
}

@article{kar2020improvement,
  title={Improvement of oral cancer screening quality and reach: The promise of artificial intelligence},
  author={Kar, Ankita and Wreesmann, Volkert B and Shwetha, Vineeth and Thakur, Shalini and Rao, Vishal US and Arakeri, Gururaj and Brennan, Peter A},
  journal={Journal of Oral Pathology \& Medicine},
  volume={49},
  number={8},
  pages={727--730},
  year={2020},
  publisher={Wiley Online Library}
}

@article{jaspers2024robustness,
  title={Robustness evaluation of deep neural networks for endoscopic image analysis: Insights and strategies},
  author={Jaspers, Tim JM and Boers, Tim GW and Kusters, Carolus HJ and Jong, Martijn R and Jukema, Jelmer B and de Groof, Albert J and Bergman, Jacques J and de With, Peter HN and van der Sommen, Fons},
  journal={Medical Image Analysis},
  volume={94},
  pages={103157},
  year={2024},
  publisher={Elsevier}
}

@article{pai2024foundation,
  title={Foundation model for cancer imaging biomarkers},
  author={Pai, Suraj and Bontempi, Dennis and Hadzic, Ibrahim and Prudente, Vasco and Soka{\v{c}}, Mateo and Chaunzwa, Tafadzwa L and Bernatz, Simon and Hosny, Ahmed and Mak, Raymond H and Birkbak, Nicolai J and others},
  journal={Nature machine intelligence},
  volume={6},
  number={3},
  pages={354--367},
  year={2024},
  publisher={Nature Publishing Group UK London}
}

@article{zuluaga2021cnn,
  title={A CNN-based methodology for breast cancer diagnosis using thermal images},
  author={Zuluaga-Gomez, Juan and Al Masry, Zeina and Benaggoune, Khaled and Meraghni, Safa and Zerhouni, Nourredine},
  journal={Computer Methods in Biomechanics and Biomedical Engineering: Imaging \& Visualization},
  volume={9},
  number={2},
  pages={131--145},
  year={2021},
  publisher={Taylor \& Francis}
}

@article{desai2021anatomization,
  title={An anatomization on breast cancer detection and diagnosis employing multi-layer perceptron neural network (MLP) and Convolutional neural network (CNN)},
  author={Desai, Meha and Shah, Manan},
  journal={Clinical eHealth},
  volume={4},
  pages={1--11},
  year={2021},
  publisher={Elsevier}
}

@article{song2021mobile,
  title={Mobile-based oral cancer classification for point-of-care screening},
  author={Song, Bofan and Sunny, Sumsum and Li, Shaobai and Gurushanth, Keerthi and Mendonca, Pramila and Mukhia, Nirza and Patrick, Sanjana and Gurudath, Shubha and Raghavan, Subhashini and Imchen, Tsusennaro and others},
  journal={Journal of biomedical optics},
  volume={26},
  number={6},
  pages={065003--065003},
  year={2021},
  publisher={Society of Photo-Optical Instrumentation Engineers}
}

@article{warin2021automatic,
  title={Automatic classification and detection of oral cancer in photographic images using deep learning algorithms},
  author={Warin, Kritsasith and Limprasert, Wasit and Suebnukarn, Siriwan and Jinaporntham, Suthin and Jantana, Patcharapon},
  journal={Journal of Oral Pathology \& Medicine},
  volume={50},
  number={9},
  pages={911--918},
  year={2021},
  publisher={Wiley Online Library}
}

@article{welikala2020automated,
  title={Automated detection and classification of oral lesions using deep learning for early detection of oral cancer},
  author={Welikala, Roshan Alex and Remagnino, Paolo and Lim, Jian Han and Chan, Chee Seng and Rajendran, Senthilmani and Kallarakkal, Thomas George and Zain, Rosnah Binti and Jayasinghe, Ruwan Duminda and Rimal, Jyotsna and Kerr, Alexander Ross and others},
  journal={Ieee Access},
  volume={8},
  pages={132677--132693},
  year={2020},
  publisher={IEEE}
}

@article{wang2022diagnostic,
  title={Diagnostic value of objective VELscope fluorescence methods in distinguishing oral cancer from oral potentially malignant disorders (OPMDs)},
  author={Wang, Caijiao and Qi, Xiangmin and Zhou, Xiaofang and Liu, Hongrui and Li, Minqi},
  journal={Translational Cancer Research},
  volume={11},
  number={6},
  pages={1603},
  year={2022}
}

@article{uthoff2018point,
  title={Point-of-care, smartphone-based, dual-modality, dual-view, oral cancer screening device with neural network classification for low-resource communities},
  author={Uthoff, Ross D and Song, Bofan and Sunny, Sumsum and Patrick, Sanjana and Suresh, Amritha and Kolur, Trupti and Keerthi, G and Spires, Oliver and Anbarani, Afarin and Wilder-Smith, Petra and others},
  journal={PloS one},
  volume={13},
  number={12},
  pages={e0207493},
  year={2018},
  publisher={Public Library of Science San Francisco, CA USA}
}

@misc{bodhade2026efficacy,
  title={Efficacy of Autofluorescence visualization devices in early detection of malignant transformation in Oral Potentially Malignant Disorders (OPMDs): A Systematic Review and Meta-Analysis},
  author={Bodhade, Ashish and Babalola, Adetola Emmanuel and Dive, Alka and Morelatto, Rosana A and Robledo, Graciela and Belardinelli, Paola and Bolesina, Nicol{\'a}s and Zapata, Marcelo and Valdez, Jesica I and Bono, Alejandra and others},
  note={Research Square preprint},
  year={2025},
  doi={10.21203/rs.3.rs-8471378/v1}
}

@article{shi2019potential,
  title={Potential role of autofluorescence imaging in determining biopsy of oral potentially malignant disorders: a large prospective diagnostic study},
  author={Shi, Linjun and Li, Chenxi and Shen, Xuemin and Zhou, Zengtong and Liu, Wei and Tang, Guoyao},
  journal={Oral Oncology},
  volume={98},
  pages={176--179},
  year={2019},
  publisher={Elsevier}
}

@article{bhokare2025diagnostic,
  title={Diagnostic Accuracy of Tissue Autofluorescence for Oral Potentially Malignant Disorders: A Systematic Review and Meta-analysis},
  author={Bhokare, Akash G and Mhapuskar, Amit and Hiremutt, Darshan R Prasad and Jadhav, Rutuja and Rao, Prashant and Kalyanpur, Kedarnath},
  journal={Journal of the International Clinical Dental Research Organization},
  volume={17},
  number={2},
  pages={117--127},
  year={2025},
  publisher={Medknow}
}

@article{wong2019using,
  title={Using photography to explore psychological distress in patients with pancreatic cancer and their caregivers: a qualitative study},
  author={Wong, Shan S and George Jr, Thomas J and Godfrey, Melyssa and Le, Jennifer and Pereira, Deidre B},
  journal={Supportive Care in Cancer},
  volume={27},
  number={1},
  pages={321--328},
  year={2019},
  publisher={Springer}
}

@article{you2025real,
  title={Real-World Readiness: Evaluating AI Model Performance in Degraded Photographic Imaging for Cancer Detection},
  author={You, Ruilin and Chen, Jiabin and Wang, Yihan and Hong, Zhihan and Wink, Cherie and Wilder-Smith, Petra and Liang, Rongguang and Song, Bofan},
  journal={Expert Systems with Applications},
  pages={130569},
  year={2025},
  publisher={Elsevier}
}

@article{thankappan2021cost,
  title={Cost-effectiveness of oral cancer screening approaches by visual examination: Systematic review},
  author={Thankappan, Krishnakumar and Subramanian, Sujha and Balasubramanian, Deepak and Kuriakose, Moni Abraham and Sankaranarayanan, Rengaswamy and Iyer, Subramania},
  journal={Head \& Neck},
  volume={43},
  number={11},
  pages={3646--3661},
  year={2021},
  publisher={Wiley Online Library}
}

@article{warnakulasuriya2021oral,
  title={Oral cancer screening: past, present, and future},
  author={Warnakulasuriya, S and Kerr, AR},
  journal={Journal of dental research},
  volume={100},
  number={12},
  pages={1313--1320},
  year={2021},
  publisher={SAGE Publications Sage CA: Los Angeles, CA}
}

@article{speight2018oral,
  title={Oral potentially malignant disorders: risk of progression to malignancy},
  author={Speight, Paul M and Khurram, Syed Ali and Kujan, Omar},
  journal={Oral surgery, oral medicine, oral pathology and oral radiology},
  volume={125},
  number={6},
  pages={612--627},
  year={2018},
  publisher={Elsevier}
}

@article{warnakulasuriya2020oral,
  title={Oral potentially malignant disorders: A comprehensive review on clinical aspects and management},
  author={Warnakulasuriya, Saman},
  journal={Oral oncology},
  volume={102},
  pages={104550},
  year={2020},
  publisher={Elsevier}
}

@article{miranda2020global,
  title={Global patterns and trends in cancers of the lip, tongue and mouth},
  author={Miranda-Filho, Adalberto and Bray, Freddie},
  journal={Oral oncology},
  volume={102},
  pages={104551},
  year={2020},
  publisher={Elsevier}
}

@article{stepan2023changing,
  title={Changing epidemiology of oral cavity cancer in the United States},
  author={Stepan, Katelyn O and Mazul, Angela L and Larson, Jeffrey and Shah, Parth and Jackson, Ryan S and Pipkorn, Patrik and Kang, Stephen Y and Puram, Sidharth V},
  journal={Otolaryngology--Head and Neck Surgery},
  volume={168},
  number={4},
  pages={761--768},
  year={2023},
  publisher={Wiley Online Library}
}

@article{conway2018changing,
  title={The changing epidemiology of oral cancer: definitions, trends, and risk factors},
  author={Conway, David I and Purkayastha, M and Chestnutt, IG},
  journal={British dental journal},
  volume={225},
  number={9},
  pages={867--873},
  year={2018},
  publisher={Nature Publishing Group}
}

@article{simonyan2014very,
  title={Very deep convolutional networks for large-scale image recognition},
  author={Simonyan, Karen and Zisserman, Andrew},
  journal={arXiv preprint arXiv:1409.1556},
  year={2014}
}

@inproceedings{huang2017densely,
  title={Densely connected convolutional networks},
  author={Huang, Gao and Liu, Zhuang and Van Der Maaten, Laurens and Weinberger, Kilian Q},
  booktitle={Proceedings of the IEEE conference on computer vision and pattern recognition},
  pages={4700--4708},
  year={2017}
}

@inproceedings{tan2019efficientnet,
  title={Efficientnet: Rethinking model scaling for convolutional neural networks},
  author={Tan, Mingxing and Le, Quoc},
  booktitle={International conference on machine learning},
  pages={6105--6114},
  year={2019},
  organization={PMLR}
}

@inproceedings{he2016deep,
  title={Deep residual learning for image recognition},
  author={He, Kaiming and Zhang, Xiangyu and Ren, Shaoqing and Sun, Jian},
  booktitle={Proceedings of the IEEE conference on computer vision and pattern recognition},
  pages={770--778},
  year={2016}
}

@article{bray2024global,
  title={Global cancer statistics 2022: GLOBOCAN estimates of incidence and mortality worldwide for 36 cancers in 185 countries},
  author={Bray, Freddie and Laversanne, Mathieu and Sung, Hyuna and Ferlay, Jacques and Siegel, Rebecca L and Soerjomataram, Isabelle and Jemal, Ahmedin},
  journal={CA: a cancer journal for clinicians},
  volume={74},
  number={3},
  pages={229--263},
  year={2024},
  publisher={Wiley Online Library}
}

@article{song2018automatic,
  title={Automatic classification of dual-modalilty, smartphone-based oral dysplasia and malignancy images using deep learning},
  author={Song, Bofan and Sunny, Sumsum and Uthoff, Ross D and Patrick, Sanjana and Suresh, Amritha and Kolur, Trupti and Keerthi, G and Anbarani, Afarin and Wilder-Smith, Petra and Kuriakose, Moni Abraham and others},
  journal={Biomedical optics express},
  volume={9},
  number={11},
  pages={5318--5329},
  year={2018},
  publisher={Optical Society of America}
}

@article{hong2022high,
  title={High-precision printing of complex glass imaging optics with precondensed liquid silica resin},
  author={Hong, Zhihan and Ye, Piaoran and Loy, Douglas A and Liang, Rongguang},
  journal={Advanced Science},
  volume={9},
  number={18},
  pages={2105595},
  year={2022},
  publisher={Wiley Online Library}
}

@misc{hong2021three,
  title={Three-dimensional printing of glass micro-optics. Optica 8 (6), 904--910},
  author={Hong, ZH and Ye, PAR and Loy, DA and Liang, RG},
  year={2021}
}

@article{uthoff2019small,
  title={Small form factor, flexible, dual-modality handheld probe for smartphone-based, point-of-care oral and oropharyngeal cancer screening},
  author={Uthoff, Ross D and Song, Bofan and Sunny, Sumsum and Patrick, Sanjana and Suresh, Amritha and Kolur, Trupti and Gurushanth, Keerthi and Wooten, Kimberly and Gupta, Vishal and Platek, Mary E and others},
  journal={Journal of biomedical optics},
  volume={24},
  number={10},
  pages={106003--106003},
  year={2019},
  publisher={Society of Photo-Optical Instrumentation Engineers}
}

@article{birur2022field,
  title={Field validation of deep learning based Point-of-Care device for early detection of oral malignant and potentially malignant disorders},
  author={Birur N, Praveen and Song, Bofan and Sunny, Sumsum P and Mendonca, Pramila and Mukhia, Nirza and Li, Shaobai and Patrick, Sanjana and AR, Subhashini and Imchen, Tsusennaro and Leivon, Shirley T and others},
  journal={Scientific Reports},
  volume={12},
  number={1},
  pages={14283},
  year={2022},
  publisher={Nature Publishing Group UK London}
}

@inproceedings{radford2021learning,
  title={Learning transferable visual models from natural language supervision},
  author={Radford, Alec and Kim, Jong Wook and Hallacy, Chris and Ramesh, Aditya and Goh, Gabriel and Agarwal, Sandhini and Sastry, Girish and Askell, Amanda and Mishkin, Pamela and Clark, Jack and others},
  booktitle={International conference on machine learning},
  pages={8748--8763},
  year={2021},
  organization={PmLR}
}

@article{hong2025dual,
  title={Dual-head multi-photon polymerization 3D printing for parallel additive manufacturing organic/inorganic materials in optics},
  author={Hong, Zhihan and Zhang, Zheng and You, Ruilin and Chen, Jiabin and Li, Shaobai and Wang, Yihan and Sun, Yuanyuan and Song, Bofan and Ji, Zhongying and Loy, Douglas A and others},
  journal={Additive Manufacturing},
  volume={103},
  pages={104772},
  year={2025},
  publisher={Elsevier}
}

@article{you2025extremely,
  title={Extremely Compact 3D Printed Glass Ternary Diffractive Optical Element for Holographic Images},
  author={You, Ruilin and Hong, Zhihan and Chen, Jiabin and Zhang, Zheng and Wang, Yihan and Sun, Yuanyuan and Song, Bofan and Loy, Douglas A and Ji, Zhongying and Liang, Rongguang},
  journal={Advanced Optical Materials},
  pages={2501074},
  year={2025},
  publisher={Wiley Online Library}
}

@article{song2025integrating,
  title={Integrating artificial intelligence with smartphone-based imaging for cancer detection in vivo},
  author={Song, Bofan and Liang, Rongguang},
  journal={Biosensors and Bioelectronics},
  volume={271},
  pages={116982},
  year={2025},
  publisher={Elsevier}
}

@article{yang2023performance,
  title={Performance of automated oral cancer screening algorithm in tobacco users vs. non-tobacco users},
  author={Yang, Susan Meishan and Song, Bofan and Wink, Cherie and Abouakl, Mary and Takesh, Thair and Hurlbutt, Michelle and Dinica, Dana and Davis, Amber and Liang, Rongguang and Wilder-Smith, Petra},
  journal={Applied Sciences},
  volume={13},
  number={5},
  pages={3370},
  year={2023},
  publisher={MDPI}
}

@article{maron2021benchmark,
  title={A benchmark for neural network robustness in skin cancer classification},
  author={Maron, Roman C and Schlager, Justin G and Haggenm{\"u}ller, Sarah and von Kalle, Christof and Utikal, Jochen S and Meier, Friedegund and Gellrich, Frank F and Hobelsberger, Sarah and Hauschild, Axel and French, Lars and others},
  journal={European Journal of Cancer},
  volume={155},
  pages={191--199},
  year={2021},
  publisher={Elsevier}
}

\end{document}